# Normative Moral Pluralism for AI:
# A Framework for Deliberation in Complex Moral Contexts

**David Doron Yaacov**

The University of Haifa

[David.d.yaacov@gmail.com](mailto:David.d.yaacov@gmail.com)

**Abstract**

The conceptual framework proposed in this paper centers on the development of a deliberative moral reasoning system—one designed to process complex moral situations by generating, filtering, and weighing normative arguments drawn from diverse ethical perspectives. While the framework is rooted in Machine Ethics, it also makes a substantive contribution to Value Alignment by outlining a system architecture that links structured moral reasoning to action under time constraints. Grounded in normative moral pluralism, this system is not constructed to imitate behavior but built on reason-sensitive deliberation over structured moral content in a transparent and principled manner. Beyond its role as a deliberative system, it also serves as the conceptual foundation for a novel two-level architecture: functioning as a moral reasoning teacher envisioned to train faster models that support real-time responsiveness without reproducing the full structure of deliberative reasoning. Together, the deliberative and intuitive components are designed to enable both deep reflection and responsive action. A key design feature is the dual-hybrid structure: a universal layer that defines a moral threshold through top-down and bottom-up learning, and a local layer that learns to weigh competing considerations in context while integrating culturally specific normative content, so long as it remains within the universal threshold. By extending the notion of moral complexity to include not only conflicting beliefs, but also multifactorial dilemmas, multiple stakeholders, and the integration of non-moral considerations, the framework aims to support morally grounded decision-making in realistic, high-stakes contexts.

## Introduction

Over the past two decades, Machine Ethics and AI Alignment have developed along largely independent trajectories. Machine Ethics, rooted in moral theory and often developed by philosophers and engineers, has focused on guiding the ethical behavior of artificial systems. In contrast, AI Alignment emerged from technical safety concerns, aiming to ensure that artificial systems behave in line with human intentions. While this alignment effort initially emphasized behavioral conformity – relying on preference learning, corrigibility, or post hoc fine-tuning – it largely avoided engagement with normative theory. Recently, however, Value Alignment, a subfield of AI Alignment, has become the site of growing convergence with Machine Ethics. As researchers confront the limitations of purely behavioral approaches, it is increasingly recognized that reliable alignment requires engagement with deeper moral content (e.g., Arnold et al. 2017; Gabriel 2020). This shift has brought core questions from Machine Ethics, such as how to integrate normative principles into AI systems and ensure their actions are justifiable across diverse moral perspectives, into the center of Value Alignment research.

The conceptual framework proposed in this paper centers on the development of a deliberative moral reasoning system—one designed to process complex moral situations by generating, filtering, and weighing normative arguments drawn from diverse ethical perspectives. While the framework is rooted in Machine Ethics, it also offers architectural insights relevant to Value Alignment by outlining how structured moral reasoning might connect to action under time constraints. Grounded in normative moral pluralism, this system is not constructed to imitate behavior but built on reason-sensitive deliberation over structured moral content in a transparent and principled manner. Beyond its role as a deliberative system, it also serves as the basis for a layered architecture: functioning as a moral reasoning teacher envisioned to train faster models that support real-time responsiveness without reproducing the full structure of deliberative reasoning. Together, the deliberative and intuitive components form a complementary structure: principled reasoning informs responsive decision-making, enabling the system to support both real-time action and deep moral reflection when circumstances allow. This dual design supports both general-purpose deployment, where reasoning and responsiveness must coexist, and domain-specific training for systems that operate within more narrowly defined ethical environments. In each case, the goal remains the same: to produce actions that are grounded in structured moral deliberation, whether directly or through learned approximation.

The deliberative system is structured as dual-hybrid architecture, combining universal normative constraints with culturally grounded flexibility. In the universal dimension, the system learns structured moral content through a combination of top-down theoretical input (moral theories, principles, and values) and bottom-up exposure to failure cases, such as cases where adherence to a principle or theory leads to outcomes that are clearly indefensible on moral grounds. Together with the rejection of relativism, this training process defines a moral threshold: a standard that excludes impermissible arguments, assumptions, and positions while preserving leeway for plural, context-sensitive moral deliberation. In the local dimension, the system is designed to incorporate additional normative material – such as cultural or institutional priorities – provided they are compatible with the universal threshold. Within these constraints, it learns to assign weights to competing considerations from the bottom up, enabling deliberation that is both principled and contextually sensitive. This structure preserves stable moral constraints while supporting morally grounded normative variation across cultures and contexts.

The remainder of the paper proceeds as follows. Section 1 examines the concept of moral complexity, extending it beyond conflicting beliefs to include multifactorial dilemmas, multiple stakeholders, and the integration of non-moral practical constraints. Section 2 outlines core limitations in Machine Ethics and two representative Value Alignment models. Section 3 introduces Normative Moral Pluralism, grounding it in philosophical theory and explaining its core structure and commitments. Section 4 presents the conceptual framework, including its dual-hybrid architecture and layered design. Section 5 outlines the deliberative process itself, detailing how the system constructs and expands moral maps, explores potential resolutions, and ultimately weighs contributory reasons to reach reasoned and context-sensitive decisions.

## 1. Moral Complexity

Moral complexity is often discussed either as a feature of moral decision-making or as a property of morality itself. This paper ultimately proposes a normative framework for ethical deliberation but begins by examining the nature of moral complexity to clarify the structural demands of serious moral reasoning. Most philosophical accounts of moral complexity focus on one central aspect: the plurality of moral values and beliefs. In Patterns of Moral Complexity (1987), Charles Larmore explores the coexistence of conflicting yet reasonable moral beliefs. He explains that moral reasoning is not simply about applying fixed rules; it requires weighing diverse values that cannot always be fully reconciled. This plurality is one of the key reasons why moral disagreements persist—not merely due to ignorance regarding empirical facts or cognitive bias, but because agents may reasonably prioritize different, yet legitimate, moral considerations. Even in the absence of explicit disagreement, practical decision-making frequently involves competing moral considerations that resist full integration, such as tensions between fairness and overall welfare in public health or infrastructure planning. This understanding of moral complexity is also expressed in parts of the Machine Ethics and Value Alignment literature, where researchers often acknowledge the diversity of moral beliefs and the persistence of disagreement. However, some approaches in these fields frame such disagreement primarily as uncertainty over which single moral theory is correct (e.g., Bogosian 2017). By contrast, the framework proposed here begins from a different premise: that moral complexity reflects the structure of moral reality itself, not merely epistemic limitations.

Beyond the plurality of moral values and beliefs, moral complexity must also be understood in a broader and more comprehensive sense, expanding the narrower focus that has dominated previous accounts. First, outside idealized theoretical contexts, moral decisions often involve practical or amoral pragmatic considerations, such as resource constraints, institutional frameworks, and the need for administratively manageable solutions to compensatory justice claims. Second, many dilemmas are structurally multifactorial, involving numerous morally significant variables and affecting multiple individuals, communities, and ecosystems simultaneously. Importantly, these factors are interconnected: resolving one dimension of a dilemma may ripple outward, altering the moral weight or feasibility of other decisions within the same context. To illustrate this, consider a public decision such as selecting the location of a new international airport. The decision impacts various stakeholders and environmental factors, including individuals, local communities, ecosystems, and future generations. Evaluating such a decision requires weighing multiple criteria: environmental impact, public health implications, economic benefits, accessibility and connectivity, land use compatibility, and long-term sustainability. Each of these criteria carries moral weight in ways that profoundly affect the lives of individual people. Their interdependencies often preclude resolution through a single evaluative perspective or straightforward trade-offs.

Such complex decisions demand multi-dimensional reasoning that integrates and balances diverse moral considerations across interconnected domains, rather than treating moral sub-questions as independent problems. Importantly, while we humans can deliberate across some dimensions of such complexity, cognitive and organizational limitations often prevent us from fully reasoning through the intricate ripple effects and trade-offs involved. Moreover, when compensatory mechanisms are employed – such as in determining financial settlements for those affected by major infrastructure projects – their solutions are often blunt, statistical, and detached from the moral textures of individual lives. This too is a form of moral loss: the erosion of fine-grained moral attention under the pressures of scale and the operational demands of administrative systems. There emerges a distinctive role for artificial moral reasoning systems: not merely replicating human moral thinking, but enabling structured, scalable moral analysis that can engage layers of interdependent moral considerations with a depth and precision that human agents, working alone, cannot feasibly achieve at scale. A framework capable of representing and navigating this broader conception of moral complexity is therefore essential for advancing the ethical capacities of intelligent systems.

On this basis, in what follows, I adopt the view that morality—and the complexity it embodies—is real, pervasive, and fundamental. Unlike human agents, artificial systems can, at least in principle, process vast amounts of morally relevant information with greater consistency and scope. When applied to domains such as public infrastructure planning, this capacity opens the door to more inclusive and morally responsive deliberation—provided the systems themselves are designed to recognize and reason through moral complexity rather than simplify it away. Such capability, whether applied in advisory roles, autonomous decision-making, or generative reasoning, demands a framework capable of engaging this complexity in practice. The normative framework developed in this paper is intended to meet that demand. Before presenting it, however, a brief overview of existing approaches in Machine Ethics and Value Alignment will help clarify how current models fall short of addressing the kind of moral complexity that real-world ethical reasoning requires.

## 2. The Limitations of Contemporary Moral Frameworks

The sense of moral complexity outlined in Section 1 is not addressed by any existing model in Machine Ethics or Value Alignment. This is understandable, given that most systems were designed for real-time applications, where decisions must be made under tight temporal constraints rather than for deliberative moral tasks such as complex planning, evaluation, or advisory contexts. Nonetheless, recent developments in Machine Ethics and Value Alignment increasingly acknowledge that AI systems must engage with a core feature of moral complexity: the diversity of moral perspectives (see, e.g., Arnold & Scheutz 2016; Bogosian 2017; Gordon 2020; Cave et al. 2019; Gabriel 2020; Martinho et al. 2021; Dobbe, Gilbert, and Mintz 2021; Song & Yeung 2024; Henschke & Arora 2024; Zhi-Xuan et al. 2024; Dubey, Dailisan, and Mahajan 2025; Tennant et al. 2025). Yet despite this growing recognition, systems designed to incorporate moral pluralism in Machine Ethics collapse into monism, while those that manage to accommodate pluralism in Value Alignment fall short on moral grounds. Rather than offering a comprehensive review of Machine Ethics or Value Alignment literature, this section outlines core limitations in classic Machine Ethics approaches, which persist in their contemporary LLM-based implementations. I then turn to two recent Value Alignment models – each exemplifying a different approach – to illustrate their underlying normative shortcomings.

Machine Ethics began with two parallel strands: top-down models based on monistic ethical theories and bottom-up models focused on behavioral imitation. Top-down models encode a single ethical theory, typically versions of deontology or utilitarianism, as a set of rules or principles applied deductively to moral situations (see, e.g., Allen, Varner, and Zinser 2000; Cloos 2005; Bringsjord, Arkoudas & Bello 2006; Powers 2006; Winfield, Blum & Liu 2014; Scheutz, Malle & Briggs 2015). In contrast, bottom-up models attempt to replicate human moral judgments by learning patterns from observed decisions without reference to explicit normative theory (see, e.g., Guarini 2006; Awad et al. 2018; Liu et al. 2022; Biltekoff 2023; Kabir et al. 2025). Top-down monistic models have been widely criticized for their rigidity, as they typically implement ethical theories as fixed sets of rules, limiting their adaptability to novel or unforeseen scenarios. They also overlook morally significant perspectives that fall outside the scope of the selected theory. Bottom-up models have been criticized for lacking transparent justification, which leads to ethically shallow outcomes. They often reflect public moral conceptions that may encode social biases, while disregarding centuries of philosophical inquiry on the nature of ethics and lacking a principled moral structure. Some hybrid models – combining symbolic structures with learning mechanisms but lacking explicit ethical theory (e.g., Wallach, Franklin & Allen 2010) – have also been criticized for normative ambiguity, inheriting the limitations of both top-down rigidity and bottom-up opacity without offering a coherent moral framework.

Some Machine Ethics models attempt to go beyond monism by incorporating multiple ethical theories or principles (for top-down models, see, e.g., Verheij 2016; Berreby et al. 2017; Zhou et al. 2023; for hybrid models, see, e.g., Anderson, Anderson & Armen 2006; Dehghani et al. 2008; Song & Yeung 2024). These models typically analyze a given situation from several ethical standpoints, aiming to identify the most appropriate course of action. However, in all these cases, the pluralistic system collapses into monism by applying one moral perspective per decision (e.g., deontology), typically by employing a meta-level criterion or a value-based ranking to determine which one is selected to guide the decision in the particular situation. Nevertheless, even in relatively common scenarios, such as caring for elderly individuals experiencing cognitive decline, morally aligned decisions often require integrating multiple ethical perspectives.

Consider an elder-care robot responding to an acute episode in which Mr. Johnson, a resident suffering from severe cognitive impairment, physically resists essential medical treatment, inadvertently endangering himself and others nearby. A consequentialist perspective suggests immediate sedation to minimize harm, yet risks psychological damage and loss of dignity. A deontological perspective cautions against violating autonomy, yet acknowledges the immediate necessity to intervene for safety. From the standpoint of virtue ethics, maintaining empathy and trust precludes harsh restraint or excessive sedation. Through real-time deliberation, the robot adopts a nuanced, multi-step response: minimal sedation solely to alleviate acute distress, gentle physical guidance without forceful restraint, and empathetic reassurance. Crucially, such an ethically preferable action emerges only from deliberation across multiple ethical perspectives, emerging from mutual adjustment and compromise among perspectives. However, because these semi-pluralistic models reduce deliberation to a single dominant perspective, they cannot support the kind of integrative reasoning required for morally nuanced decisions. As a result, they fall short of aligning with the complexity of human ethical reasoning.

Unlike the models discussed so far, recent approaches in Value Alignment do not collapse pluralism by selecting a single ethical theory at decision time. Instead, they often preserve multiple perspectives side by side, attempting to capture user-aligned moral preferences by combining outputs from different ethical viewpoints. One such model, proposed by Dognin et al. (2024), offers a method for responding to moral scenarios by drawing on several ethical perspectives at once. For each situation, the system produces a set of short moral statements, each expressing what a different ethical viewpoint recommends (for example, "Calm Mr. Johnson down" from a care perspective). These statements are then merged into a single response. The way they are merged depends on how much weight the system learned to assign each moral perspective based on similar contexts during training.

While this kind of model may work well in relatively simple situations – such as issuing a directive like "Gently calm Mr. Johnson with minimal sedation, administered slowly, and provide clear, reassuring verbal cues such as 'You're safe here,' in a soft and steady tone," which reflects multiple ethical perspectives – it faces serious limitations in contexts involving genuine moral dilemmas. In such dilemmas, all available options are ethically troubling, each supported and opposed by different moral perspectives. Here, we are not just navigating between competing values but between incommensurable wrongs. Thus, when time permits, we have a moral obligation to engage in profound reasoning to uncover relevant arguments, reexamine overlooked features, and refine our grasp of the relevant stakes. Ethical judgment requires deliberation, which can often involve surfacing considerations that alter our original assessment of the context. This is consistent with findings from moral case deliberation in healthcare, where structured discussion has been shown to uncover previously unnoticed considerations and incorporate fine-grained empirical details into ethical assessment (see e.g., Molewijk et al. 2019). Once our understanding of the context shifts, the weights we initially assigned will no longer apply. A system that rigidly applies pre-learned contextual weights risks moral failure—not due to incorrect scoring, but because it misunderstood what it was scoring in the first place. Moral alignment is not just about expressing human preferences; it is also about deliberating reflectively when the situation demands it. Without this capacity, the system risks failing in its core purpose: to align with human moral judgment, which often depends on rethinking the situation through deeper deliberation.

Another recent example of a pluralist framework in Value Alignment is Dubey et al. (2025). In their model, the system begins by mapping out a full space of possible actions to complete a given task. Each action is then evaluated from five predefined distinct moral perspectives, with the system assigning a numerical score and a brief textual explanation for each. Once all actions are scored, the model aggregates each action's scores across perspectives. The selected action is the one that best completes the task while receiving the highest combined moral score—reflecting either maximum agreement or minimum conflict among the different ethical viewpoints.

However, as Dubey et al. explain, their base agents are trained to pursue predefined goals, and the moral scoring only shapes how those goals are achieved—not whether they should be achieved at all. This reveals a deeper issue that extends beyond their model. Much of the Value Alignment literature equates alignment with human preferences to moral adequacy, as if the problem were simply to follow what people want (Zhi-Xuan et al. 2024). But not all human desires are ethically justified, and not all assigned tasks are morally acceptable. A truly moral system must be able to reason about the permissibility of its objectives—not only the means to reach them. Otherwise, it risks implementing unethical goals through superficially aligned actions. My concern extends further: Even when the goal itself is morally praiseworthy, it may be that every available action carries a significant moral cost. For example, we may want to launch a weather balloon to collect valuable data, but not if every launch site available requires destroying a village's only source of clean water. Every system must be able to assess whether both its goals and the means of achieving them are morally permissible. This assessment may be represented mathematically, but it must result from structured and intelligible moral reasoning—reasoning carried out in natural language, with clear deliberative steps we can understand and evaluate.

Aggregation-based approaches can handle the narrow sense of moral complexity better than current Machine Ethics models (including recent LLM-driven ones). However, they flatten the moral landscape and cannot substitute for deep moral evaluation when such reasoning is required. These limitations underscore the need for systems capable of genuine moral reasoning. Normative moral pluralism offers a compelling foundation for this task—one that supports deliberation across diverse perspectives and engages with complexity.

## 3. Normative Moral Pluralism

### 3.1. What is Normative Moral Pluralism?

According to normative moral pluralism, in many instances, different and even conflicting normative positions (e.g., deontological, consequentialist, and egalitarian) can all be morally acceptable in a single case. This viewpoint acknowledges that diverse moral views, theories, and principles contribute essential dimensions to the moral deliberation process that a single view, theory, or principle cannot encompass. As a normative framework, it strives to incorporate a wide range of ethical perspectives, including various moral principles, theories, and intuitions, not limited to philosophical literature, such as religious traditions, cultural practices, or professional codes of conduct.

Normative moral pluralism operates on two levels. First, it incorporates pluralism of considerations by integrating diverse ethical perspectives into the reasoning process. Second, it supports pluralism of decisions by recognizing that multiple morally reasonable outcomes may exist. Moral pluralism stands in clear opposition to moral relativism. While some advanced versions of relativism appeal to cultural coherence or internal justification rather than unrestricted moral validity, they still fall short of providing criteria for rejecting oppressive or harmful norms across contexts. By contrast, normative moral pluralism maintains both a respect for diversity and a commitment to universal moral boundaries. Thus, one of the central features of its normative framework is the concept of a moral threshold, which excludes morally unreasonable options while allowing for a range of morally acceptable courses of action. These boundaries are not arbitrary, and their philosophical justification and specific nature will be elaborated in the next sub-section.

Pluralism of considerations involves simultaneously assessing multiple moral reasons—such as those stemming from duties, consequences, or diverse moral values. To genuinely weigh these different reasons, a pluralistic framework must treat them explicitly as contributory rather than decisive. On this view, moral reasons are not overriding imperatives that automatically determine action; instead, each reason contributes partial weight, enabling context-sensitive, flexible deliberation. While a single contributory reason can, in some cases, carry overwhelming weight and decisively support one course of action, the structure remains pluralist, grounded in weighing rather than rule application. Treating moral considerations as contributory reasons is thus an essential feature of pluralism of considerations, as it naturally supports resolutions that integrate or balance competing values.

Unlike monistic frameworks, pluralism of considerations enables creative solutions that integrate diverse ethical principles, thereby allowing a richer and more structurally responsive approach to complex moral reasoning. These solutions may consist of multiple steps or concurrent actions, each shaped by distinct moral considerations, resulting in outcomes that express a more complex and nuanced ethical position. For instance, a morally creative resolution in infrastructure planning might involve rerouting part of a project to reduce environmental harm while offering targeted support to affected communities—thus integrating concerns of sustainability, fairness, and access. While such ethical creativity is not a conceptual requirement of normative moral pluralism, it is inherently supported by the treatment of reasons as contributory.

Though not required by the framework, credence-based reasoning offers a practical and effective way to operationalize this flexibility, since credences can capture varying degrees of weight attributed to diverse moral considerations. In this context, credences are not treated merely as measures of epistemic confidence, but as proxies for the contributory force of normative considerations—representing the degree to which each should figure into moral deliberation. This makes it particularly well suited for addressing moral complexity, especially in deliberation over complex decisions involving numerous variables and affected stakeholders, such as those encountered in public infrastructure planning.

### 3.2. The Philosophical Foundations of Normative Moral Pluralism

Many philosophers have endorsed pluralistic positions regarding morality, albeit in varying forms. The normative framework I propose here draws substantially on the tradition of value pluralism—a view that has received considerable philosophical support and has shaped key debates about moral complexity. Some of the most influential formulations include Berlin (1958), Strawson (1961), Larmore

(1987), and Kekes (1996). They reject the idea that moral conflicts can always be resolved by appealing to a single overarching value, such as the claim that promoting overall well-being should consistently take precedence over values like justice, integrity, or loyalty. Instead, they emphasize the reality of deep, sometimes tragic, value conflicts that reflect irreducible features of moral life.

Berlin's seminal work (1958) provides a significant foundation for this view. His account of value pluralism centres on the idea that fundamental human values—such as liberty, equality, loyalty, and justice—are all objectively significant, yet often incompatible and incommensurable. These values may conflict in ways that cannot be resolved by appealing to a common standard or higher-ranking value, since no common metric exists for comparing them. Consequently, these incommensurable values are often viewed as equally fundamental, each holding its own intrinsic worth without a clear hierarchical ordering. While Berlin offers no prescriptive method for navigating conflicting values, his recognition of this structure of moral reality supports the idea that moral reasoning must remain open to multiple perspectives and forms of justification.

Berlin, along with those who follow him in expanding the pluralist tradition, consistently rejected moral relativism. Despite acknowledging the plurality and conflict of values, they remain committed to the view that morality is objective and that moral values have genuine normative force. Strawson (1961) adds a crucial distinction between universal values—those that apply to all moral agents—and more localized values, which reflect justified variations across cultures or communities. This distinction will become important later when we consider how AI systems might draw on normative pluralism to reason about moral questions across different contexts and cultures.

While Berlin and Strawson lay the philosophical groundwork by analysing moral conflicts and value plurality, Rawls (1971) offers a methodological instantiation of pluralism of considerations through reflective equilibrium, providing a systematic approach to integrating a diverse range of moral intuitions, principles, and judgments. Although Rawls does not explicitly use the term contributory reasons, his method implicitly treats moral considerations as contributory—each consideration provides partial support rather than being an absolute determinant of moral decisions. This idea, explicitly articulated later by Jonathan Dancy (2004, especially Chapter 2), highlights moral reasons as context-dependent and variable in weight, reinforcing the alignment between pluralism of considerations and moral contextualism. Indeed, both reflective equilibrium and Dancy's particularism emphasize sensitivity to context, recognizing that the moral significance of considerations can shift depending on the situation. Thus, pluralism of considerations — and, by extension, normative moral pluralism — naturally fits within a broader contextualist understanding of ethical reasoning.

Rawls's reflective equilibrium allows for ethical creativity by enabling the mutual adjustment of principles, intuitions, and considered judgments. It permits the reinterpretation and integration of diverse moral considerations in the process of arriving at a coherent moral judgment. A similar openness to ethical creativity also appears in Smilansky's work (2019, forthcoming, especially Chapters 1 and 2), as he accepts the blending of conflicting reasons and values in ethical deliberation. There, Smilansky offers a clear instance of pluralism of both considerations and decisions—the recognition that multiple morally reasonable reasons, decisions, and outcomes may coexist within a bounded moral space. He demonstrates that rejecting moral pluralism as a normative approach is untenable in resource allocation dilemmas, effectively establishing that moral pluralism is the only viable framework for such cases. Smilansky's rejection of moral relativism, along with his critique of decision-making based on immoral, arbitrary, or morally ungrounded criteria, inherently employs the concept of a moral threshold to exclude unacceptable options. A related concern with delineating limits within value pluralism, though developed in a different context, also appears in Kekes (1996, especially Chapter 2).

In previous work (Yaacov 2022), I argued that moral pluralism provides a robust framework for addressing a wide range of moral dilemmas, including those often viewed as resistant to pluralistic interpretations. I proposed that by employing nuanced credence-based reasoning — treating degrees of confidence in normative propositions as proxies for contributory weight — one can identify multiple, contextually supported resolutions within a bounded moral space, even in cases typically assumed to demand a single correct moral response. Although we often have a preferred course of action in a given dilemma, we may see other options as morally reasonable if they are supported by serious considerations. In situations where all available actions involve moral cost, our choices are frequently made without full conviction, not because of epistemic weakness, but because competing options retain genuine moral force. The choice, therefore, is not a matter of epistemic uncertainty or ignorance, but a matter of weighing morally significant considerations where no option fully dominates the others. Judgment, in these cases, is not a knockout but a decision on points, reflecting a comprehensive understanding of the moral landscape rather than a failure to grasp it. I further contended that while moral pluralism accommodates diverse perspectives, its moral threshold must exclude extreme options, such as refusing to sacrifice one person to save a million others in a trolley-type scenario, even if supported by deontology, a widely recognized central ethical theory—since adhering to it in such a context lies outside the boundaries of morality due to its catastrophic consequences.

While I do believe that pluralism of values, considerations, and decisions offers the most accurate representation of the complex nature of ethics, its practical value remains even for those who maintain that, under ideal conditions, a single morally correct resolution exists for each dilemma (see, e.g., Rawls). Given the non-ideal conditions under which both humans and intelligent systems must operate — such as limited information, resource constraints, or time pressure — normative moral pluralism offers the most suitable framework for enabling intelligent systems to navigate complex moral landscapes. With this groundwork in place, the next step is to examine how such a framework might be effectively implemented in intelligent systems.

## 4. The Deliberative Framework

### 4.1. Establishing the Framework

Building on the normative framework developed in Section 3, this part of the paper outlines a conceptual blueprint for a system-level implementation framework that enables intelligent machines to deliberate over complex moral situations in a principled yet context-sensitive manner. Rather than relying on rule-based programming or behavior imitation, the system operates entirely through LLMs, which learn moral content from language, reason through language, and express decisions in language. The system follows a dual-hybrid approach operating along two dimensions: universal and local. At the universal level, it ensures that machines can distinguish between morally acceptable and unacceptable forms of reasoning and action. At the local level, it learns weights and enables cultural and institutional adaptation within those boundaries.

The universal level defines the foundational structure of the system's moral reasoning, combining moral knowledge and constraints that define a threshold for admissible deliberation. The moral threshold functions across three distinct phases: it constrains local adaptation by filtering out morally invalid learning, shapes internal deliberation by excluding inadmissible arguments, and protects against unacceptable external input during third-party engagement. Establishing it requires addressing three challenges. First, the system must exclude decisions based on arbitrary or morally ungrounded criteria. Second, it must exclude unjustified reasoning and clearly indefensible outcomes. Third, it must prevent the inclusion of norms that cannot be justified beyond their local or subjective origins, thereby avoiding relativism. To implement the threshold, the system must first undergo top-down training with structured moral knowledge—philosophical literature on ethics that satisfies the requirement of rejecting relativism. While specifying this moral corpus requires separate analysis, it can be assumed to include major theories and perspectives in the philosophical ethics literature. Since the threshold operates not as a set of action-guiding rules, but as a constraint on which arguments may enter the deliberative space, this process enables the system to engage in moral reasoning grounded in that corpus, while excluding justifications that fall outside its normative foundations.

Establishing the threshold is complemented by a targeted bottom-up process designed to identify cases where the application of otherwise admissible values, principles, or theories could lead to morally indefensible outcomes. For example, while compatriot partiality is recognized as a legitimate moral value, its unqualified application could justify discriminatory or exclusionary practices under the guise of loyalty or identity. Similarly, classical act utilitarianism, which focuses on maximizing overall utility, might justify harvesting organs from one person to save five—a conclusion that contradicts widely held moral intuitions and illustrates where the theory overextends its appropriate scope. To prevent this, the system is trained to recognize such cases and to apply these values or principles in a constrained and context-sensitive manner. Crucially, it also learns to generalize—identifying structurally similar patterns in new contexts where the application of this otherwise admissible moral element leads, once again, to unjustified outcomes. Such generalization cannot be reliably achieved through top-down programming or fine-tuning alone.

Once the universal moral threshold is established, the system enters a local adaptation phase, aligning with community values while ensuring these remain within universal moral boundaries. Like the threshold itself, this process is hybrid. Through top-down guidance, locally relevant ethical content—such as values derived from cultural traditions, religious texts, or professional codes of conduct—may be introduced, provided it meets the conditions of the universal moral threshold. In addition, stakeholders may specify which of these values to prioritize, so long as those priorities do not conflict with universal constraints. Any top-down preferences that conflict with the threshold are excluded from the adaptation process. For example, a country may prioritize meritocracy—valuing excellence and rewarding individual achievement—even at the expense of egalitarian principles. Both meritocracy and egalitarianism fall within the system's moral boundaries, as they reflect differing value priorities without violating ethical constraints. In contrast, preferences that advocate discrimination against marginalized groups, such as religious minorities, would be rejected for lacking moral justification.

This top-down process is complemented by bottom-up learning. The primary role of the local bottom-up phase is to learn how different moral considerations are weighted across varying contexts within a given community. It does so by identifying patterns in moral choices and adjusting the relative weight assigned to different values. All learned adaptations—such as context-sensitive weightings or emphasis shifts—remain within the bounds set by the universal level. For instance, people in low-crime societies may prioritize privacy over security, while those in high-crime areas may accept some infringements on privacy as a trade-off for safety. These are valid contextual preferences that fall within the accepted moral framework. By contrast, a claim that undocumented immigrants are not entitled to basic rights or protections would be excluded, as it lacks a defensible moral basis. Determining whose moral judgments should guide this learning process—whether those of experts, laypersons, or some combination—is itself a substantive question (see e.g., Riesen and Boespflug 2025), but it should be assumed that only those judgments consistent with the universal threshold can shape the system's normative calibration.

Together, the universal threshold and local adaptation define the boundaries and texture of the system's moral landscape, establishing the foundation on which deliberative reasoning can proceed. In practice, this means that similar cases arising in different cultures may lead to different yet morally acceptable outcomes, each shaped by local priorities but constrained by the same underlying moral standard. Having outlined the structure that enables deliberation within a bounded moral space, the next step is to consider how such reasoning can support timely action in applied settings.

### 4.2. From Deliberation to Fast Moral Action

Teacher–student architectures are already well established in machine learning, particularly in knowledge distillation and imitation learning (see, e.g., Hu et al. 2023; Messikommer et al. 2025). These approaches typically involve training a smaller or faster model (the student) to replicate the outputs of a larger, more complex model (the teacher), allowing the student to approximate the teacher's behavior with reduced computational cost. Such architectures carry promise both in contexts where moral decisions must be carried out in real time, and in shaping system behavior in ways that reflect background social values, similar to the role of etiquette, when no explicit moral judgment is required.

In a full deployment, the framework is designed as a layered system in which a deliberative model and an intuitive model work in tandem. The deliberative component serves as the normative expert: it constructs a moral map of the situation, filters inadmissible reasoning, and produces justified decisions. The intuitive model approximates these outputs, enabling fast, low-latency behavior when time constraints prevent full deliberation. The two models are not run in parallel but serve distinct roles—reasoned output generation and efficient approximation—and together form a complete system for value-aligned action under varying practical constraints. This architecture mirrors dual-process models in cognitive psychology, which distinguish between fast, intuitive processes and slower, deliberative reasoning, each playing a complementary role in human decision-making (see, e.g., Dual-Process Theories, Encyclopedia of Social Psychology, 2007). Although the heuristic model cannot generate justifications on its own, the deliberative system that trained it retains the capacity to reconstruct the reasoning behind each decision, allowing for post-hoc explanations when needed. This setup offers a lightweight yet ethically grounded solution for systems that must act quickly without abandoning moral coherence.

While the full layered system is suitable for deployment in complex or open-ended environments, the same training approach could also support more narrowly focused systems, such as surgical triage assistants or autonomous delivery drones. In such cases, a heuristic model might be trained in advance on a set of deliberative outputs, learning how to respond in specific, well-characterized scenarios. This kind of expert training reduces the likelihood of moral error by narrowing the system's behavioral scope and limiting the need for generalization. Although the feasibility of this approach depends on the stability of the domain and the reliability of the deliberative training, it offers a promising lightweight alternative when full deliberation is neither practical nor necessary. These possibilities raise deeper design questions about how moral competence can be distributed across systems of varying complexity and purpose.

A central design question concerns what the intuitive model should be trained on. One possibility is to train it on structured moral maps generated by the deliberative system—compact representations of the moral space for a given scenario, including filtered arguments and assigned weights. Another option is to train it on representations of the world-state itself, teaching it to associate morally relevant patterns in context with appropriate behavior. It is also possible to imagine hybrid configurations where partial maps, contextual signals, or simplified ethical sketches serve as training inputs. Each option presents distinct trade-offs in terms of generalizability, interpretability, and robustness. Future work may also examine whether the deliberative system can support long-term refinement of the intuitive model, allowing for ongoing behavioral alignment over time. This would mirror the way reflective reasoning in humans sometimes guides and reshapes habitual responses.

This section has outlined a conceptual blueprint rather than a finalized system. Many design choices remain open, and the framework leaves space for further theoretical and technical refinement. The aim here is to provide a structured foundation for continued exploration—not to settle the architectural details in advance. The next section turns from architectural structure to the reasoning process itself, outlining how deliberation unfolds within the moral space defined here.

## 5. The Deliberative Process

Designing the structured reasoning process of a deliberative system raises questions that cannot be fully resolved in a single paper. In fact, it is likely that some critical issues will not even be identified here. The process proposed below is not a closed algorithmic procedure, but a preliminary design intended to guide future research and, eventually, implementation. What follows is a conceptual sketch—an outline of the deliberative process intended to reflect the core commitments of the normative framework while leaving room for refinement and debate.

The deliberative process presupposes that the system has already recognized a situation as morally significant. Yet this assumption masks a deep methodological gap: most work in Machine Ethics and Value Alignment presumes that moral salience is either pre-identified or contextually obvious. Diamond (2025) recently proposed a cognitively inspired framework that uses reflex-based triggers, grounded in psychological models of affect and survival, to initiate ethical processing. While this offers a promising technical scaffold, it does not engage with normative theory or philosophical research on moral deliberation. Still, psychologically plausible mechanisms may help systems detect ethically charged situations. Here, I proceed stipulatively: assuming that some form of salience detection has occurred, the focus shifts to the deliberative process itself.

Once a situation has been identified as morally significant, the system constructs a preliminary moral map—a sparse, structured representation of the scenario's ethically salient features, such as the relevant stakeholders and the basic contours of the potential conflict. The map serves not as a judgment but as an interpretive frame: it outlines what is morally at stake without attempting to resolve the situation. Typically, this map is built from features of the world-state, combined with structured, relevant moral knowledge. Yet even this initial step raises important questions. What counts as a sufficient map? Which features should be included, and how are morally relevant patterns distinguished from incidental ones? These questions remain open, but for present purposes, I assume that the system is able to generate a minimally adequate representation that supports further analysis, much like most contemporary work in the field, which typically presents moral conflicts in simplified form with a single central justification attached to each competing viewpoint.

Too much of the literature in Machine Ethics and Value Alignment treats every morally significant situation as a dilemma, collapsing the distinction between ordinary conflict and genuine moral tragedy. This flattening of the moral landscape obscures crucial structural differences. As the familiar problem of moral dilemmas in the technical philosophical sense makes clear (see e.g., Foot 1983, Statman 1995), not all conflicts are dilemmas. A true dilemma arises only when all available actions involve morally significant loss, and no resolution can preserve all legitimate claims (Statman 1995). In the context of AI, this distinction is not merely theoretical—it shapes how systems ought to deliberate. Misclassifying a conflict as a dilemma may lead the system to expend unnecessary effort, while treating a genuine dilemma as a resolvable conflict risks moral blindness, as discussed in Section Two.

Three main strategies can resolve moral conflicts without escalating them into full dilemmas, as Statman explains: integration, where a creative solution sufficiently addresses all competing demands; compromise, where conflicting claims are partially fulfilled; and compensation, where one claim is set aside but its moral cost is addressed through a reparative act that fully offsets the loss and eliminates the conflict. These methods reflect the capacity of a deliberative system to dissolve conflict without incurring moral loss. When such resolutions are unavailable, the system must proceed to a deliberative override—assessing the competing contributory reasons and selecting the action supported by the strongest overall set. Even when one option is ultimately chosen, the deliberative process does not end there. As Statman notes, the moral significance of the defeated option does not disappear; it may generate new obligations, such as an apology or compensation. These second-order demands become part of the system's ongoing moral reasoning, allowing it to register the normative cost of action and respond appropriately. Statman's framework provides a principled basis for distinguishing between these cases and enables the deliberative process to respond accordingly.

Deliberation begins with the assumption that the moral conflict might be resolvable through integration, compromise, or compensation—strategies grounded respectively in creativity, negotiated concession, and compensatory or reparative ethics. Rather than selecting from predefined options, the system reasons from admissible moral content, generating and comparing arguments to determine justified actions. Initially, the system attempts to creatively resolve the conflict by exploring potential solutions based on the preliminary moral map and relevant normative theory—for instance, by achieving integration, as when a robo-nurse places a quiet fan near a patient who wants a breeze, avoiding the need to open a window that would disturb another patient sensitive to cold. As it evaluates these solutions, it raises targeted questions whose answers gradually enrich the moral map by testing empirical assumptions, verifying argumentative validity, expanding contextual information, and incorporating additional morally relevant considerations. This iterative refinement leverages the generative ca-pabilities of LLM-based inference models, allowing the system to question its assumptions, generate diverse and context-sensitive solutions, and propose resolutions reflecting contributory reasons and learned normative weights.

If no satisfactory resolution emerges from this preliminary deliberation, the system escalates to deeper pluralist reasoning. At this stage, the system maximally expands its moral map, exploring additional normative and empirical dimensions to fully understand the dilemma. This includes not only the contributory reasons already mapped, but also second-order reasons related to the moral weight of creatively crafted reparative or compensatory responses. In such cases, the potential for compensation does not eliminate the harm, but may shape how the system compares inescapable losses. For example, if both outcomes involve serious irreversible injury, but one action leads to the collapse of a person's identity, vocation, or life structure, while the other allows for some recovery or continuity, this difference becomes a morally relevant factor. Rather than treating all harm as equal, the system considers the meaning and consequences of what remains. The result is a reason-sensitive judgment that reflects the normative force of all arguments—including those that recognize the limits of repair.

The deliberative process described here is not intended as a finalized design, but as a conceptual sketch that outlines how such reasoning might be structured in practice. Many open questions remain. How should the system determine when its moral map is sufficiently expanded to proceed from conflict resolution to dilemma management? What forms of prompting are best suited to elicit this kind of reasoning from large language models? How should the system prioritize among conflicting reasons when context-sensitive weights are close or contested? These and other questions lie beyond the scope of the current paper. Still, they help clarify the remaining work. By framing deliberation as a structured, pluralist, and reason-sensitive process—rather than as static preference modeling or rule execution—the framework aims to support the development of systems that can respond to moral complexity in a principled and explainable manner.

# 6. Conclusion

As AI systems increasingly engage with morally significant decisions, foundational questions about the structure of ethical reasoning have become urgent. Although Machine Ethics and Value Alignment developed along largely separate trajectories, recent work has brought them into closer contact, as researchers increasingly acknowledge that reliable alignment requires engagement with deeper moral content. This paper responds to that convergence by proposing a structured deliberative framework grounded in Normative Moral Pluralism—a philosophically rigorous approach designed to inform the development of intelligent systems capable of reason-sensitive alignment.

Unlike approaches that treat moral complexity as a constraint to be managed through heuristics, value encoding, or optimization, the framework developed here begins from the assumption that such complexity must be addressed directly through structured moral reasoning. This framework meets that demand through dual-hybrid architecture: a universal layer that sets a normative moral threshold, and a local layer that allows for cultural adaptation within those constraints. Built on this foundation, the system's deliberative process is implemented through LLM-based inference, enabling it to construct moral maps, expand relevant considerations, and weigh contributory reasons. This mode of deliberation supports a form of creative reasoning—allowing the system to explore alternative framings, recognize second-order considerations, and generate context-sensitive responses to both ordinary conflicts and genuine moral dilemmas.

Another key innovation is the system's layered structure, which pairs a deliberative process with an intuitive, inference-speed model. The deliberative system generates outputs that serve as training data for the faster model, enabling real-time moral responsiveness without sacrificing normative depth. This design mirrors dual-process models in human cognition, allowing the system to act swiftly while remaining grounded in explainable, value-sensitive reasoning.

Recent work has highlighted the challenge of calibrating aligned systems to act with the appropriate degree of moral concern in context (Firt 2023). While the conceptual framework developed here does not resolve the calibration problem, it offers a necessary foundation for addressing it: a deliberative process that produces structured, language-based moral reasoning, capable of expressing both ethical justification and normative constraint. Whether such reasoning can reliably shape behavior remains an open question—one that is likely to require additional mechanisms beyond language alone. But by making the system's ethical reasoning explicit, inspectable, and grounded in principled pluralism, this framework shifts the problem from one of intuition to one of implementation. In doing so, it strengthens the need for engineering, and, perhaps even psychological, solutions that help ensure intelligent moral agents follow the moral logic they are capable of articulating.